\documentclass[iop,useAMS,usenatbib]{emulateapj}
\usepackage{epsfig}
\usepackage{natbib}
\usepackage{subfigure}
\usepackage{amssymb,amsmath}

\DeclareMathVersion{bold}
\newcommand{\eg}{e.\,g.}
\newcommand{\dmunits}{{\rm pc\,cm^{-3}}}

\newcommand{\src}{FRB\,011025}

\newcommand{\ti}{t_{\rm i}}
\newcommand{\tI}{t_{\rm I}}
\newcommand{\taus}{\tau_{\rm s}}

\newcommand{\tsamp}{t_{\rm samp}}

\newcommand{\dmmw}{\rm DM_{MW}}
\newcommand{\Nt}{N_{\rm T}}
\newcommand{\Na}{N_{\rm A}}
\newcommand{\ra}{\rightarrow}
\shorttitle{\src and the Galactic Position Dependence of FRBs}
\shortauthors{S.~Burke-Spolaor \& K.~W.~Bannister}

\begin{document} 
\title{The Galactic Position Dependence of Fast Radio Bursts and the Discovery of \src}

\author{Sarah Burke-Spolaor}
\affil{California Institute of Technology, 1200 E California Blvd, Pasadena CA, USA}
\affil{Jet Propulsion Laboratory, California Institute of Technology, Pasadena CA, USA}
\email{sarahbspolaor@gmail.com}
\and
\author{Keith W.~Bannister\altaffilmark{1}}
\affil{CSIRO Astronomy and Space Sciences, P.O. Box 76, Epping NSW 1710, Australia}
\altaffiltext1{Bolton Fellow}

\begin{abstract}
We report the detection of a dispersed Fast Radio Burst (FRB) in archival intermediate-latitude Parkes Radio Telescope data. The burst appears to be of the same physical origin as the four purported extragalactic FRBs reported by \citet{thornton13}. This burst's arrival time precedes the Thornton et al.~bursts by ten years.
We consider that this survey, and many other archival low-latitude ($|gb|<30^\circ$) pulsar surveys, have been searched for FRBs but produced fewer detections than the comparatively brief Thornton et al.~search.
Such a rate dependence on Galactic position 
could provide critical supporting evidence for an extragalactic origin for FRBs. To test this, we form an analytic expression to account for Galactic position and survey setup in FRB rate predictions. Employing a sky temperature, scattering, and dispersion model of the Milky Way, we compute the expected number of FRBs if they are isotropically distributed on the sky w.r.t.\ Galactic position (i.\,e.~local), and if they are of extragalactic origin. We demonstrate that
the relative detection rates reject a local origin with a confidence of 99.96\% ($\sim$3.6$\sigma$). The extragalactic predictions provide a better agreement, however are still strong discrepancies with the low-latitude detection rate at a confidence of 99.69\% ($\sim$2.9$\sigma$). However, for the extragalactic population, the differences in predicted vs.~detected population may be accounted for by a number of factors, which we discuss.
\end{abstract}

\section{Introduction}\label{sec:intro}

The High Time Resolution Universe (HTRU) survey, currently underway at the Parkes telescope \citep{htru1}, recently uncovered what appears to be an extragalactic population of single-bursting events \citep{thornton13}. Such events have previously been theorized to arise from a number of cataclysmic astrophysical events such as black hole evaporation \citep{rees77} and neutron star coalescence \citep{lipaczynski98,hansenlyutikov01}.
The four Fast Radio Bursts (FRBs) published by \citet{thornton13} have so far represented the clearest evidence of such a population, followed by the additional recent discovery at Arecibo Telescope by \citet{spitler14}.
If these FRBs are extragalactic, they hold vast 
potential
as accurate probes of ionized intergalactic media, and to study the exotic physical conditions that cause them. Unfortunately, the isolated millisecond-duration pulses detected by single-dish telescopes have poor sky localization, with positional errors of up to 30\,arcminutes.

Interferometric radio experiments will eventually provide localized FRB detections to precisely isolate burst origins. Meanwhile, \emph{any} FRB discoveries will contribute to the distribution statistics of this potentially cosmological population. Understanding FRB distributions in dispersion, flux, and sky position are critical to make key assessments of likely FRB origins. Recently, the analysis of \citet{petroff14} indicated that there appears to be a dearth of detections in the HTRU survey intermediate latitude regions.

In this paper, we report a new FRB detection in archival Parkes Radio Telescope data, and provide a robust method to assess the relative detection rates in a range of archival surveys at low Galactic latitude. We apply this method to a number of archival surveys to arrive at a confident statement of a Galactic latitude dependence of FRBs. Throughout this analysis, we take the following precautions with two previously purported FRBs:
\begin{enumerate}
\item The \citet{LB} detection: This burst's properties can be explained by both the terrestrial ``Peryton'' phenomenon of \citet{perytons}, and the FRBs of \citet{thornton13}. Furthermore, this burst was discovered in a survey of the Magellanic Clouds, where there are significant errors in the Galactic electron density model used in our analysis \citep{ne2001}. To avoid introducing unwelcome biases from these ambiguities, we entirely exclude \emph{both} the burst and the survey that gave rise to this detection from our analysis of Galactic position dependence of FRBs. This decision makes our study robust against future revelations that the Lorimer et al.\ detection is a member of either phenomenon.
\item The \citet{keane} detection: This burst's dispersion measure had only a small excess over that expected from a standard Galactic source, \eg\ a neutron star. The report of \citet{bannister14} has shown by analysis of Galactic ionized hydrogen in the Keane burst line-of-sight that this burst has a high probability of being a Galactic object. We thus do not consider this burst in our analysis, however do employ its survey data.
\end{enumerate}

\section{Data \& Analysis}\label{sec:data}
We inspected the archival surveys of \citet{ED} and \citet{BJ}, which collectively included Galactic latitudes $5^\circ<|gb|<30^\circ$ at longitudes $-100^\circ< \ell < 50^\circ$. The survey configurations were identical, sampling a 96-channel 288\,MHz bandwidth centred at 1372.5\,MHz. The time sampling for polarization-summed power samples was 125\,$\mu$s, and each pointing lasted 4.4\,minutes.

The original single pulse processing of these surveys was reported in \citet{sbsmb}.
We reprocessed the surveys using the search and automatic classification procedures of \citet{htruSP}
which, in brief, involved seeking events over a signal-to-noise ratio (S/N) threshold of 6 in a range of dispersion measure and boxcar filter trials. Our boxcars were of size $2^n$, ranging $0\leq n\leq9$. Candidate clusters are identified as events using a friend-of-friends method; single-member events are rejected as Gaussian noise, and events with a maximum signal-to-noise ratio at ${\rm DM}\leq1.5\,\dmunits$ are automatically rejected as local interference.
In addition to this process, to raise the sensitivity to bursts like those reported by \citet{thornton13}, we:
\begin{itemize}
\item Searched DM trials up to 3000\,$\dmunits$, higher than the previous search limit of 600\,$\dmunits$.
\item Employed pre-processing interference excision based on the coincidence filter techniques of \citet{kocz12} to minimize the false positive rate of burst candidates.
\end{itemize}
All events ranked as valid candidates were assessed manually using the same inspection plots and methods as \citet{htruSP}. However, only pointing summary plots (not individual event plots) were inspected for pointings in which all candidates were of ${\rm DM}\leq 200\,\dmunits$. As with other experiments of this kind, we estimate that candidates could be reliably identified down to ${\rm S/N}\simeq7$ (previous inspections of single pulses in a series of candidates from rotating radio transients have demonstrated that manual plot inspection is unreliable for ${\rm S/N}<7$; \eg\ \citealt{htruSP}).

\section{Properties of FRB\,011025}
\begin{figure}
\centering
\includegraphics[width=0.47\textwidth,trim=0mm 0mm 0mm 3mm, clip]{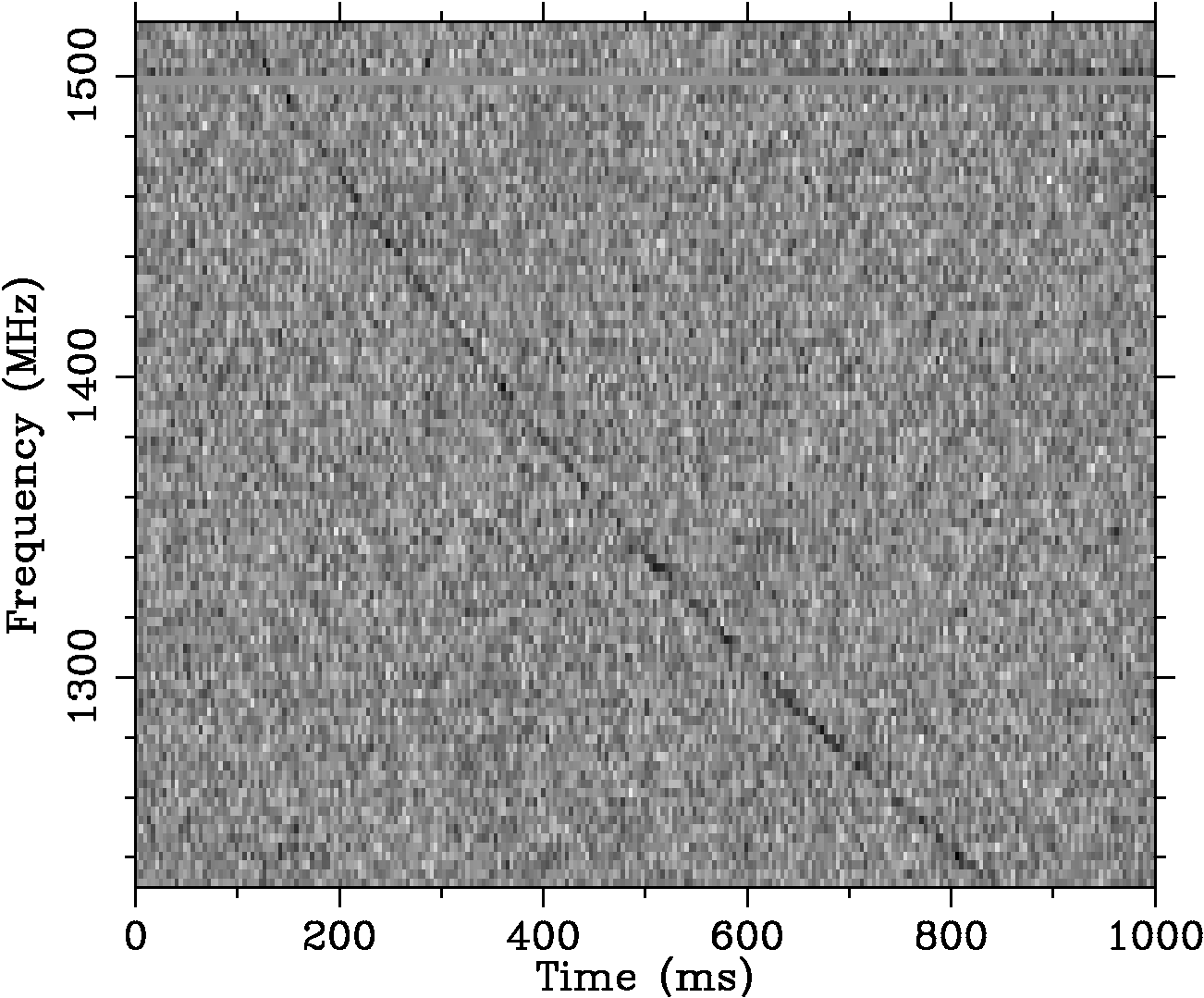}
\caption{The spectrogram of \src, with arbitrary power scale plotted in reverse greyscale. One frequency channel containing known narrow-band interference ($f\sim1500$\,MHz) was removed.
}\label{fig:waterfall}
\end{figure}

Only one burst with a DM exceeding the expected Galactic contribution, and not clearly associated with interference, was discovered in our search. The burst was detected at a signal-to-noise ratio of 16.9. We report the observed properties of this burst in Table\,\ref{table:params}, and the data spectrogram is shown in Fig.\,\ref{fig:waterfall}.

\begin{figure}
\centering
\includegraphics[width=0.47\textwidth,trim=24mm 29mm 19mm 22mm, clip]{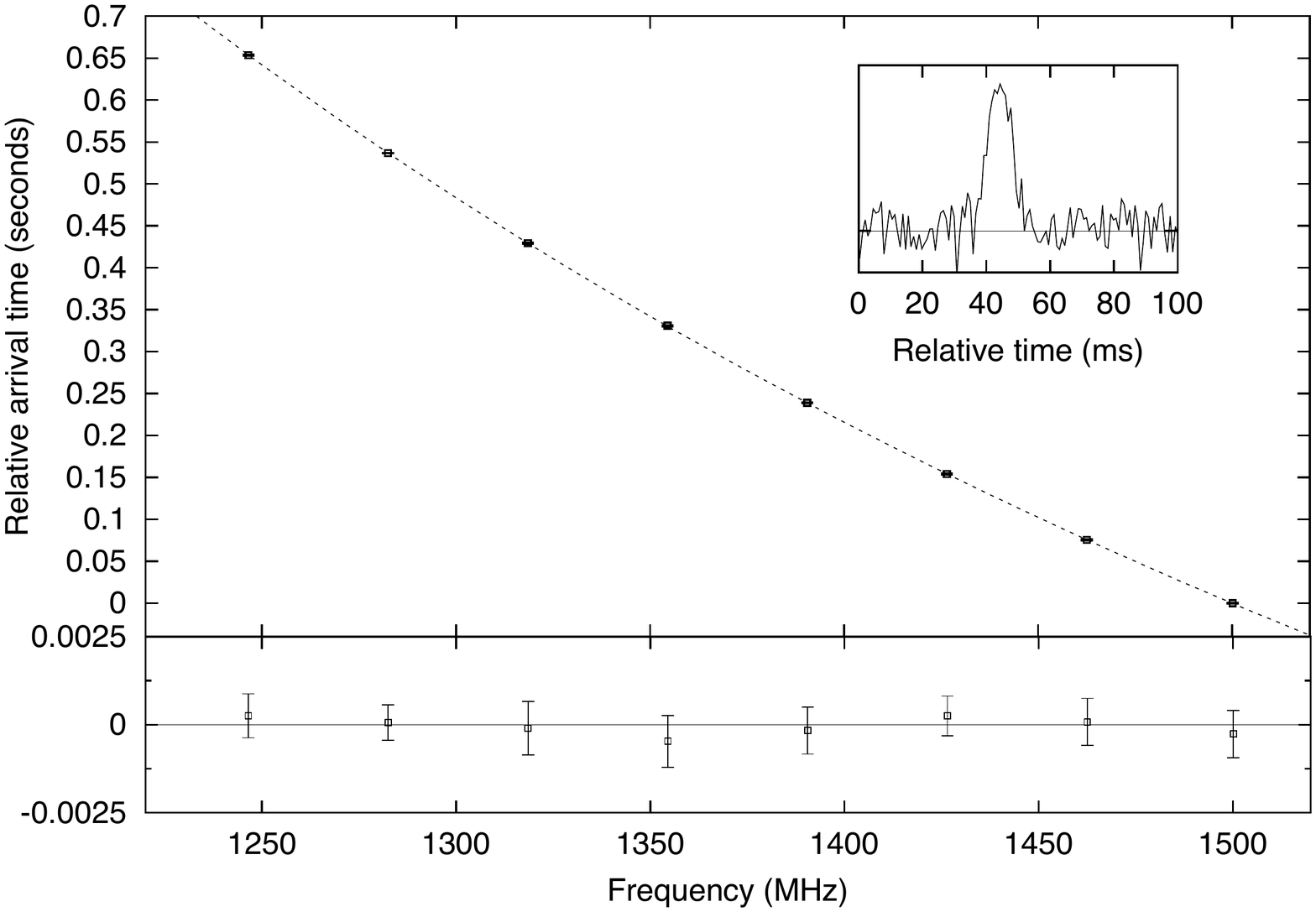}
\caption{The frequency-dependent delay of \src\ in eight sub-bands. Prior to delay measurement, each sub-band was corrected for a dispersive delay with \mbox{${\rm DM}=790\,\dmunits$}. An analytic template was cross-correlated with that from each sub-band to determine its temporal delay as per standard pulsar timing procedure \citep[e.\,g.][]{psrchive}.$^1$ The dotted line shows a least squares fit with free parameters DM and $a$, where $dt \propto {\rm DM}\cdot f^{a}$. The results of this fit are reported in Table\,\ref{table:params}. The inset figure shows the integrated pulse shape after correction at ${\rm DM}=790\,\dmunits$.
}\label{fig:delays}
\end{figure}

This FRB has multiple properties that closely resemble the bursts reported by \citet{thornton13}. First, the frequency-dependent delay closely follows a cold plasma dispersion relation, indicative of propagation through astrophysical plasmas (Fig.\,\ref{fig:delays}), and unlike the terrestrial-origin ``perytons''. The observed DM over that expected to be contributed by our Galaxy is a factor of $\sim$7 based on the {\sc ne2001} model's predicted dispersion at this line-of-sight, giving a large DM excess, ${\rm DM_E}$, and indicative of an extragalactic origin for the burst. Assuming the fully-ionized IGM model of \citet{inoue04} and attributing all of ${\rm DM_E}$ to dispersion in the IGM, we get an upper limit on redshift for our burst of $z\leq 0.69$. For comparison with the redshifts reported by \citet{thornton13}, assuming a 100\,$\dmunits$ contribution from a galactic host indicates a redshift $z=0.59$. This implies a maximum co-moving distance to the target of $D\leq 2.4\,$Gpc, and $D_{100}=2.1$\,Gpc, respectively.

The burst has a duration of $w_{50}=9.4$\,ms at its 50\% power point. The width of the pulse is largely contributed by intra-channel dispersion smearing (7.9\,ms; see Eq.\,\ref{eq:ti} below). using Using Eq.\,\ref{eq:ti} to accounting for intra-channel smearing and the sample time, we find a 5\,ms pulse width after removing instrumental broadening.

To test whether the pulse shows evidence for frequency dependent scatter-broadening, as the brightest of the Thornton et al. bursts did (they found $w_{50}\propto f^{\mu}$, where $\mu=-4.0\pm0.4$), we measured the pulse width in four independent sub-bands, and found that the pulse is broadened at low frequency with a best-fit index of $\mu=-4.2\pm1.2$, consistent with scatter broadening and with the index of the Thornton burst. Our pulse does not show significant asymmetry or a pulse tail (the light curve's skewness measured $\pm$35\,ms around the pulse peak is -0.018), however noise in the data could easily be masking a scattering tail. The inferred pulse width of \src\ rivals that of the longest-duration FRB\,110220, which measured 5.6\,ms and exhibited a clear scattering tail. The shorter durations of other FRB detections and our pulse broadening measurement imply that this pulse is likewise scatter-broadened.

Based on the similarity of their basic observed properties (excess dispersion, duration, fluence [time-integrated flux]), the burst appears to be of the same population as those reported by \citet{thornton13}. We will assume that these bursts are all drawn from the same physical phenomenon until other evidence motivates a re-examination of this assumption.

\begin{table}
\begin{centering}
  \caption{Properties of \src. Parameter $t_{1516.5}$ refers to the arrival time of the burst at 1516.5\,MHz.}\label{table:params}

 \begin{tabular}{rl}
\hline
Pointing RA, Dec (J2000) & 19:06:53.0,  -40:37:14.4\\
Pointing $gl, gb$ & 356.641, -20.021\\
$t_{1516.5}$\,(UTC Y-M-D, h:m:s) & 2001-01-25, 00:29:$13.23\pm0.02$ \\
Delay index ($dt\propto f^{a}$) & $-2.00\pm0.01$\\
DM& 790 $\pm$ 3$\,\dmunits$\\
DM$_{\rm E}$ & 680\,$\dmunits$\\ 
$w_{\rm 50}$ & $9.4\pm0.2$\,ms\\
Scattering index ($w_{\rm 50}\propto f^{\mu})$ & $-4.2\pm1.2$\\
$S_{\rm peak, min}$ & 300\,mJy\\ 
Fluence & $<2.82\times10^{-3}$\,Jy\,s\\
\hline
\end{tabular}
\end{centering}
\end{table}



\section{Burst Rates versus Galactic Latitude}
A contrast in our discovery versus the \citet{thornton13} burst discoveries is the difference in detected rate in our survey. Our archival intermediate-latitude data covered a total of 11858.76 hours on the sky, counting each 0.043\,deg$^2$ beam in the Multibeam receiver as an independent data stream. This implies a detection rate in the survey of \mbox{$\sim$$2\times10^3{\rm \,sky^{-1}\,day^{-1}}$}, a factor of $\sim$5 smaller than the reported HTRU high-latitude detection rate. This gap widens when considering the non-detection of FRBs in other low-latitude surveys, notably the PKSMB survey of \citet{pksmb} and its Perseus Arm counterpart \citep{burgay13}, the PALFA survey reported by \citet{deneva}, and the search of 23.5\% of the High Time Resolution Universe intermediate-latitude region, reported by \citet{htruSP}. We aim to examine the statistical significance and origin of the detection deficit in these lower latitude surveys.

It is not immediately obvious whether this difference in detection rate is due to differences in sky coverage, or differences in sensitivity of each survey. Therefore, below we formulate the relative detection rates in surveys of different design and sky area.


\subsection{Survey design and sky position considerations}

For a pulse of intrinsic width $\ti$ scattered to width $\taus$, the minimum flux detectable by a survey performing a standard threshold-based search is given by 
\begin{equation}\label{eq:s}
S =\frac{m\,\eta\,(T_{\rm rec}+T_{\rm sky})}{G\sqrt{n_{\rm p}\,M\,B\,\tI}} \times \frac{\tI}{\sqrt{\ti^2+{\taus}^2}}~,
\end{equation}
where $m$ represents the S/N threshold for candidate pulses; $\eta$ is an efficiency factor to account for sampling/correlator imperfections; $G$, $n_{\rm p}$ and $B$ give the telescope gain, number of polarizations and the total bandwidth, respectively; and $T_{\rm rec}$ and $T_{\rm sky}$ represent the receiver and sky temperatures, respectively. $M$ is a factor to account for dish configuration, \eg\ $M=1$ for single dish detection, or $M=n(n-1)$ for imaging with an $n$-element interferometer.
We assume that a system will integrate a pulse to time $\tI$ ideally matched to the intrinsic width and scattering, within limitations of the native sampling time $t_{\rm samp}$ and pulse broadening effects due to finite channel width $b$:
\begin{equation}\label{eq:ti}
\tI^2 = 
\bigg(
\frac{\ti}{\mu{\rm s}}\bigg)^2 + 
\bigg(\frac{\tsamp}{\mu{\rm s}}\bigg)^2 + 
\bigg(\frac{\tau_{\rm s}}{\mu{\rm s}}\bigg)^2 + 
\bigg(\frac{k\,{\rm DM}\,b}{f^3}\bigg)^2 
~\mu{\rm s}.
\end{equation}
Here, $k=8.3$, $b$ is in MHz, and $f$ is in GHz.\footnote{We assume here that the optimal integration width occurs roughly at the band center, and note that this is valid unless the burst has a very steep or inverted spectrum (which is not the case for FRBs so far). For a flat spectrum source, the band-averaged DM broadening for a flat-spectrum target across the band differs negligibly, by a fractional change of $<$2\% for all the surveys we consider.} The scattering timescale is frequency-dependent; it may scale from a reference frequency such that $\tau_{\rm s} = \tau_{\rm 0}(f/f_0)^{\mu}$. 


We now build a predictive expression based on Eq.\,\ref{eq:s} and \ref{eq:ti}, and scaling factors for instantaneous field of view $\Omega$, observation time $t_{\rm obs}$, and spectral index. The number of detections expected in a single survey pointing $p$ in survey 1, $N_1(p)$, when compared to the total number of detections, $N_2$, discovered in the entire survey 2, is:
\begin{align}
\begin{split}\label{eq:rates}
N_1(p) & = N_2
\cdot \bigg(\frac{G_2m_1\eta_1\cdot[T_{\rm rec1}+T_{\rm sky1}(p)]}
{G_1m_2\eta_2\cdot(T_{\rm rec2}+\langle T_{\rm sky2}\rangle)}\bigg)^\gamma~
\bigg(\frac{B_1M_1}{B_2M_2}\bigg)^{-\gamma/2}
\\
&\times
\bigg(\frac{t_{\rm I1}(p)}{\langle t_{\rm I2}\rangle}\bigg)^{\gamma/2}
\bigg(\frac{f_1}{f_2}\bigg)^{\alpha} 
\bigg(\frac{t_{obs}(p)}{T_2}\bigg)
\bigg(\frac{\Omega_1}{\Omega_2}\bigg)
~;\\\\
& = N_2\cdot \theta(p)
\end{split}
\end{align}
Here, the index 1 or 2 represents the respective survey, $T_{\rm sky}$ is the sky temperature in the pointing direction of the observation, and $T_2$ is the total observing time of survey 2. The exponent $\gamma$ is a power-law index representing the observed flux distribution of the population ($N\propto S^{\gamma}$). The use of a power-law index here implicitly assumes that we can model any redshift-dependent evolution of the FRB population's luminosity with a power-law.
For a non-evolving population in a Euclidean Universe, $\gamma=-3/2$; this is a reasonable assumption for FRBs for a generic $z\lesssim1$ astrophysical population.
The averages over $T_{\rm sky2}$ and $t_{\rm I2}$ represent the pointing-averaged values for that survey; in the calculation of $t_{\rm I2}$, the average values of DM and $\taus$ across the survey positions may be used if accounting for DM and $\taus$ effects.
We define spectral index as $S\propto f^\alpha$. $T_{\rm sky}$ is also frequency-dependent (it can change by a factor of two across a $\sim$300\,MHz band at 1\,GHz), however we use the center frequency of each survey in this analysis to represent the pointing's value. 

$\theta(p)$ encompasses all the expected scaling factors for pointing $p$. In our analysis below, we compare the detection rate in the archival surveys with the detection rate of \citet{thornton13}, such that ``survey 2'' uses the parameters of Thornton et al., and $N_2=\Nt=4$. The total number of events expected in the archival surveys is the sum of Eq.\,\ref{eq:rates} over every pointing in each survey:
\begin{equation}\label{eq:na}
N_A = N_{\rm T}\cdot\sum^{n_{\rm surv}}_{s=1} \sum^{n_{\rm point}}_{p=1} \theta_s(p)
 = N_{\rm T}\,\Theta
~,
\end{equation}
where we have simplified the additive scaling for all pointings in our survey set to be represented by $\Theta$.


\begin{figure*}
\centering
\subfigure[$\gamma=-1$]{
\includegraphics[height=0.175\textheight,trim=32.5mm 26.5mm 32.5mm 88mm, clip]{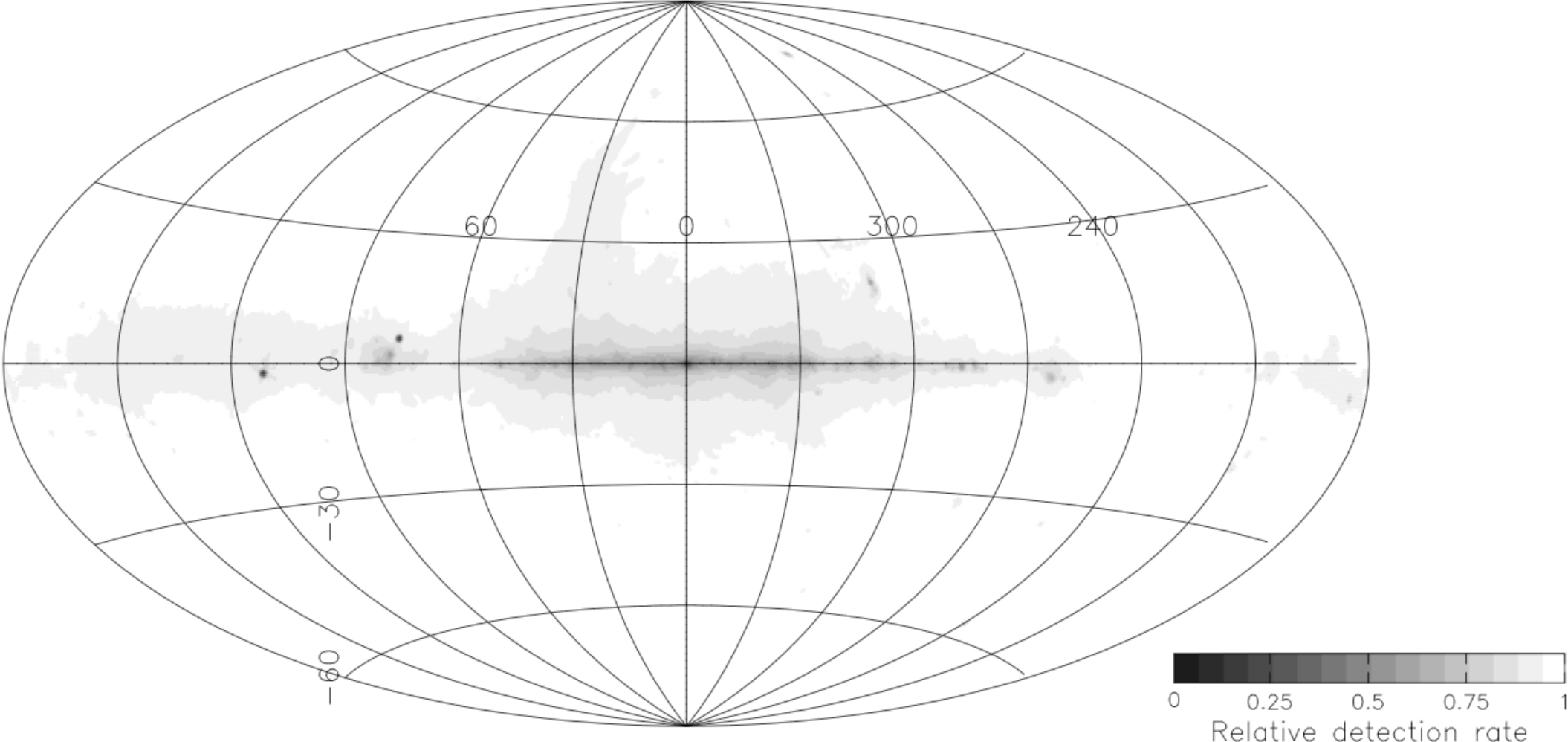}
}\quad
\subfigure[$\gamma=-2$]{
\includegraphics[height=0.175\textheight,trim=32.5mm 26.5mm 32.5mm 88mm, clip]{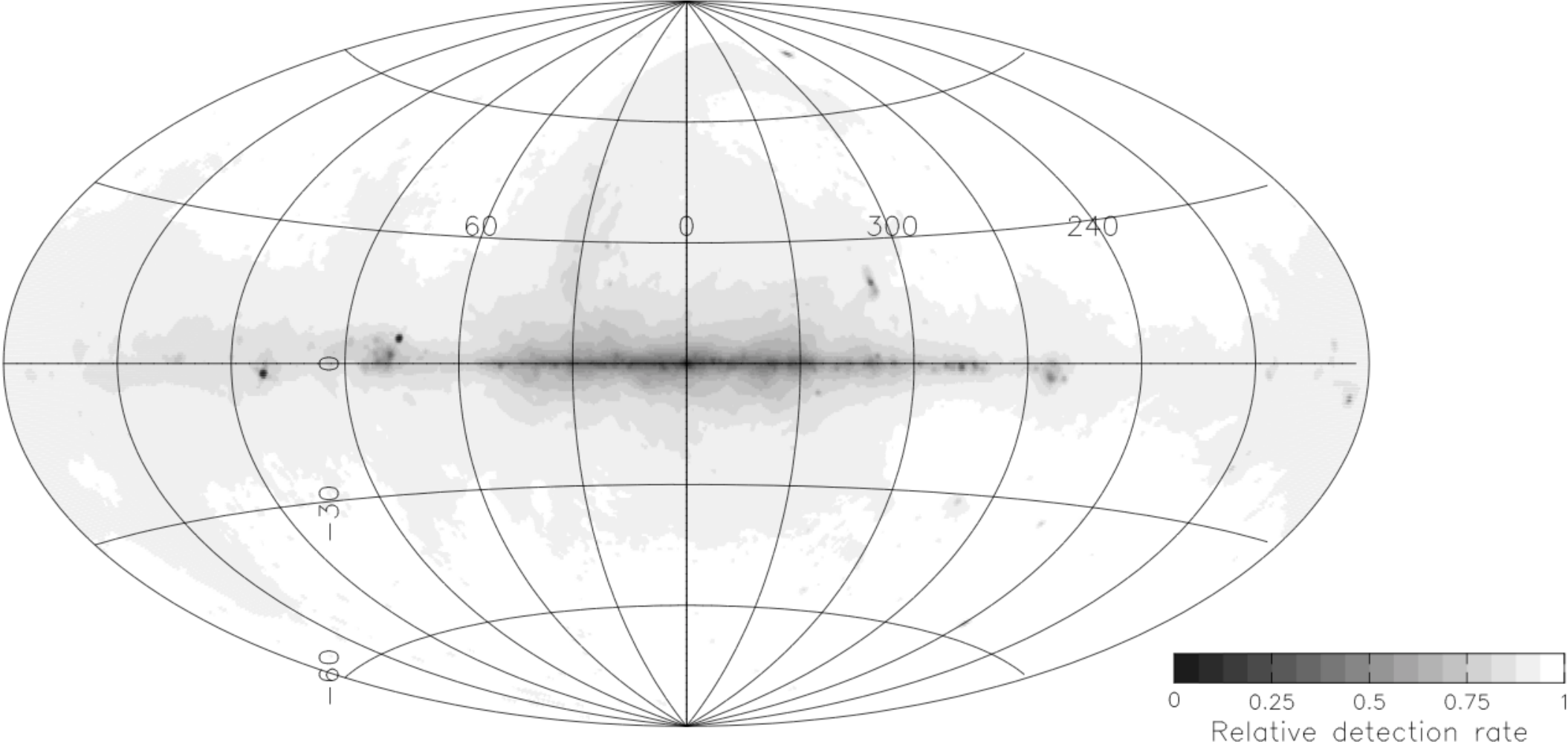}
}
\caption{Here we show the relative detection rate versus sky position for a local/isotropic population (i.\,e.~terrestrial or solar system; \S\ref{sec:local}) of FRBs using a telescope with $T_{\rm rec}=23$\,K, and the global sky temperature model of \citet{globalskymodel}. The color scale is referenced from the point on the sky with a maximum detection rate. A local population's relative rates depend only on $T_{\rm sky}$, such that the sky dependence gradient is relatively shallow, particularly for flatter pulse energy distributions.}\label{fig:local}
\end{figure*}

\begin{figure*}
\centering
\subfigure[Parkes Analogue Filterbank]{
\includegraphics[height=0.175\textheight,trim=32.5mm 26.5mm 32.5mm 88mm, clip]{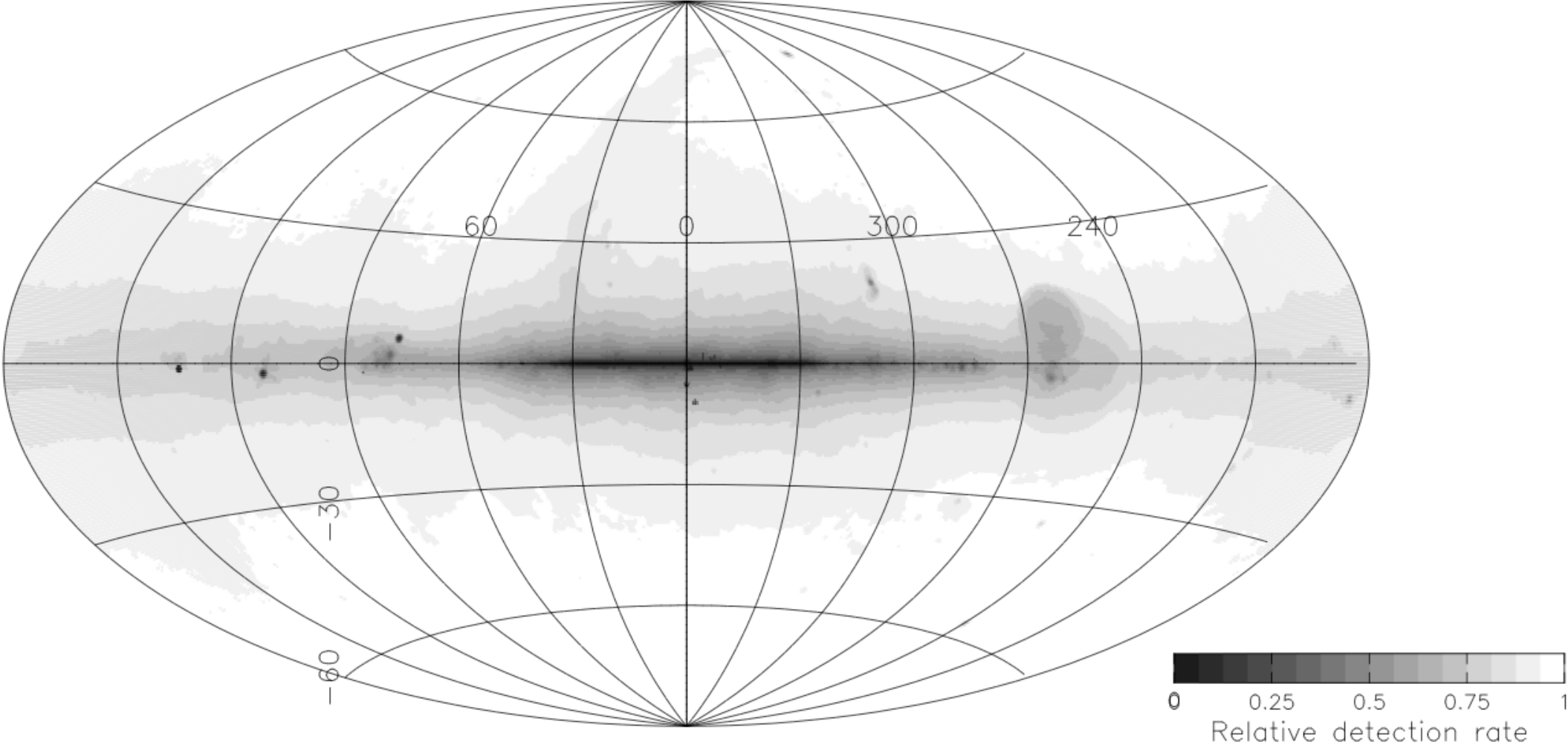}
}\quad
\subfigure[HTRU Survey]{
\includegraphics[height=0.175\textheight,trim=32.5mm 26.5mm 32.5mm 88mm, clip]{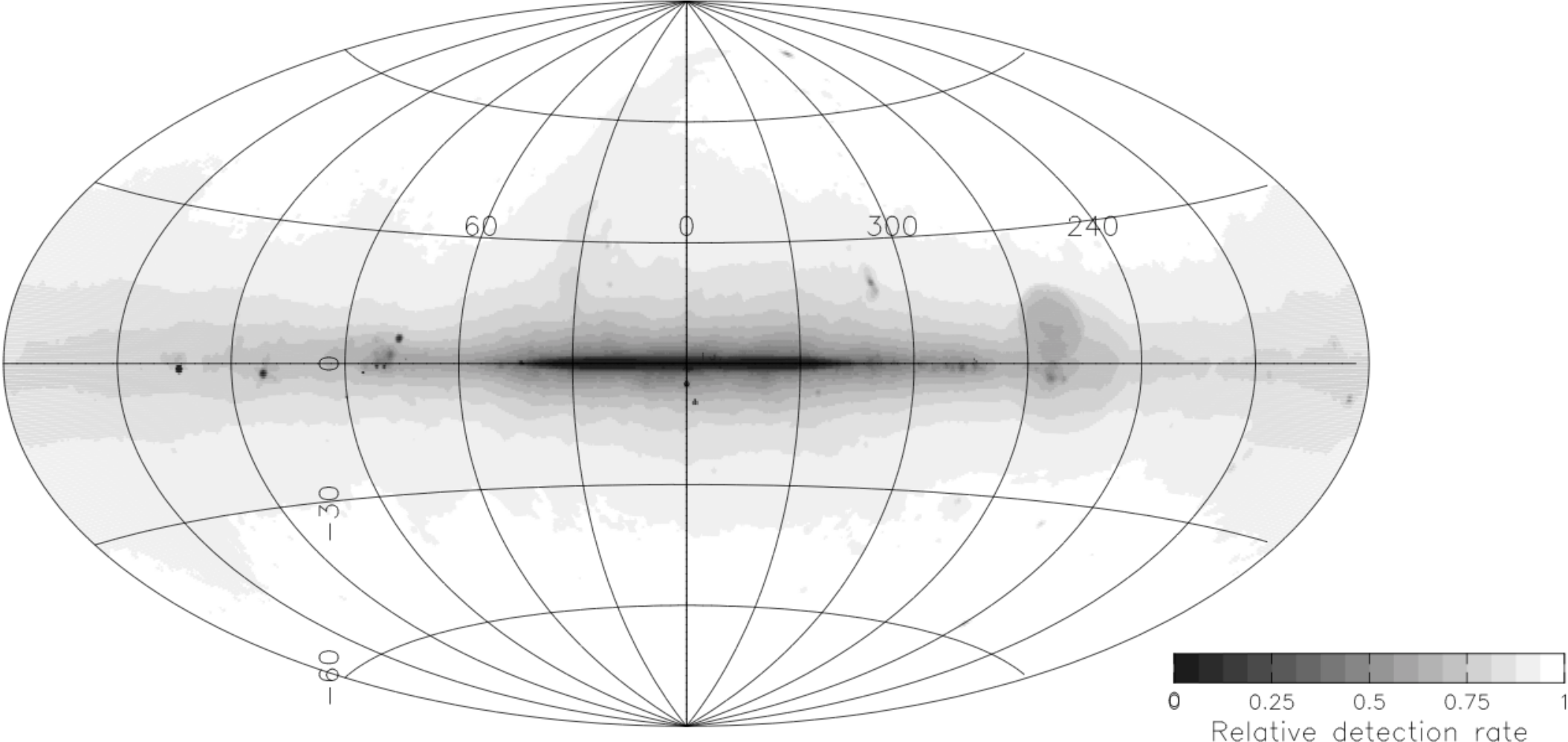}
}
\caption{As in Fig.\,\ref{fig:local}, but for an extragalactic population with $\gamma=-1.5$ using the {\sc ne2001} electron density model to predict the scattering and dispersion influence of the Milky Way. (\emph{a}) shows the dependence given the instrumental set-up for our data, and (\emph{b}) shows the dependence for the \citet{thornton13} instrumentation. These plots reflect sky-dependent scalings specific to these instruments, which performed the majority of the surveys analyzed in this paper.
A combination of dispersion, scattering, and $T_{\rm sky}$ contribute to a much steeper detection rate gradient at $|gb|<30^\circ$ than for a local population for both instruments. The higher frequency and time resolution in the Thornton et al.\ data cause their $|gb|\lesssim5$ rate to be more greatly effected by scatter broadening than the analogue filterbank.
}\label{fig:exgal}
\end{figure*}


\subsection{Inputs for ``local'' and ``cosmological'' populations}\label{sec:gpos}
We can now use the sky-dependent scalings of Eq.\,\ref{eq:rates} to demonstrate the anticipated detection rate vs.\ sky position for FRBs. With an extragalactic population, one must account for $gl$ and $gb$-dependent contributions to ${\rm DM}$, $\taus$, and $T_{\rm sky}$, which may serve to dampen the number of expected detections at decreasing $|gb|$. However, any signal not propagating through the Galaxy (i.\,e.\ bursts of instrumental, terrestrial, or even solar system origin)---hereafter we refer to these as a ``local/isotropic'' population---will have detection rate dependencies affected only by $T_{\rm sky}$, which effectively raises the receiver noise at low $|gb|$.
For the remaining analysis in this paper, we assume basic properties of the FRB population based on what has thus far been observed, using the four bursts of \citet{thornton13} and our burst.

\subsubsection{Local/isotropic population}\label{sec:local}
Here we refer to any terrestrial or solar system origin which has no intrinsic clustering with respect to Galactic position. For instance: local FRBs might arise from self-generated signals in the receiver or software path, or from an atmospheric origin (which might not be isotropic in azimuth/elevation, but which would be uncorrelated with Galactic coordinates). Aircraft or satellite origins may appear isotropic, depending on the timing and route of the flight path. Likewise, ``local'' signals from planetary origins in or out of the ecliptic plane should appear isotropically distributed with regards to Galactic position, as long as the survey being searched is well-distributed in observing time and day of year.

A detection rate dependence on sky position for a local FRB population will arise only from the scaling \mbox{$(T_{\rm rec}+T_{\rm sky})^\gamma$}.
All other factors in Eq.\,\ref{eq:rates} will not induce position dependence for a local FRB population. We compute the position-dependent $T_{\rm sky}$ at 1.4\,GHz using the multi-frequency global sky temperature model of \citet{globalskymodel}. The resulting dependence of detection rate with sky position for a local population is shown in Fig.\,\ref{fig:local} for a range of $\gamma$ values.

\subsubsection{Cosmological population}\label{sec:cosmo}
For the extragalactic FRB scenario, in addition to sky temperature, the factors that will influence position dependence of FRB detection rate are excess scattering and dispersion induced by the Milky Way's interstellar medium. These terms enter the rate through Eq.\,\ref{eq:ti}.
We take the mean \emph{extragalactic contribution} to DM for the population to be $\langle {\rm DM_E}\rangle=772\,\dmunits$.
To evaluate the DM at which FRBs will be detected, for each Galactic position we add $\langle {\rm DM_E}\rangle$ to the total Milky Way DM contribution at that position, $\dmmw$, employing the current standard {\sc ne2001} electron density model of the Galaxy \citep{ne2001} to evaluate $\dmmw$.

We approximate Galaxy-induced scattering ($\taus$) by using the {\sc ne2001} model to determine the distance, $d_{\rm sc}$, at which the Galactic scattering timescale is maximized ($\tau_{\rm max}$). That is, $\tau_{\rm max}$ is the largest scattering that can be induced in any pulse along that line of sight according to the {\sc ne2001} model; for extragalactic-origin FRBs, the scattering will always be smaller than this value.

To estimate the Galactic scattering experienced by an extragalactic FRB, we follow the analysis of \citet{lorimer13}, taking \mbox{$\taus = 4\,\tau_{\rm max}\,(1-f)f$}, where $f$ is $d_{\rm sc}$ divided by the source distance.
We thus are required to select an average detectable FRB distance with which to estimate the magnitude of Galactic scattering. As long as we select a sufficiently high value, it will not greatly impact the results of this analysis because for current instrumentation, only FRB distances below a few hundred Mpc will cause Galactic scattering to dominate Eq.\,\ref{eq:ti} for Galactic latitudes beyond $|gb|\sim3$. FRB distances have only upper limits, so we use 1\,Gpc to represent an arbitrarily large average detectable FRB distance, such that $f=d_{\rm sc}/(1$\,Gpc).
We make the assumption that all bursts have an intrinsic duration $t_{\rm i}$ much less than instrumental broadening and scattering effects, and use $\mu=-4.0$ to appropriately scale Galactic scattering \citep{bhatetal04}.
The results of this analysis for two instruments are shown in \mbox{Figure \ref{fig:exgal}}.

\begin{table*}
\caption{Survey parameters of various transient searches and results of our relative detection rate analysis. 
Surveys are: Our search (Swinburne Multibeam surveys, SWMB); Parkes Multibeam Survey (PKSMB; \citealt{keane,keanemusings}, Mickaliger et al.~in prep.), the Perseus Arm survey \citep{burgay13}, \citet{thornton13} (T13), the HTRU intermediate latitude survey (P14, \citealt{petroff14}; see also \citealt{htru1} and \citealt{htruSP}), and two PALFA surveys, which used the  7-beam 21\,cm transient search system on Arecibo (D09, \citealt{deneva} and S14, \citealt{spitler14}).
For the SWMB, T13, D09, and S14 surveys, we computed $\dmmw, T_{\rm sky}, \taus,$ and the numerical results using actual survey pointing lists (T13 pointings, D.\,Thornton/B.\,Stappers, private comm; D09, J.\,Deneva/A.\,Brazier, private comm.).
For the other surveys, where pointing lists were unavailable, we calculated these values using an evenly-spaced grid of pointings across the quoted sky area of the survey. 
}\label{table:sparams}
\centering
{\scriptsize 
\begin{tabular}{r|ccc|cc|cc|c}
\hline
&\multicolumn{3}{c|}{\textbf{Parkes Archival Surveys}} & \multicolumn{2}{c|}{\textbf{HTRU}} & \multicolumn{2}{c|}{\textbf{PALFA}} \\
 &\textbf{SWMB}& \textbf{PKSMB} & \textbf{Perseus Arm} & \textbf{T13}& \textbf{P14} & \textbf{D09} & \textbf{S14} \\
\hline
$gb$ range &$5^\circ<|gb|<30^\circ$ & $|gb|<5^\circ$ & $|gb|<5^\circ$& All-sky, & $|gb|<15^\circ$ & $|gb|\leq5^\circ$& $|gb|\leq5^\circ$\\
$gl$ range&$-100^\circ\ra50^\circ$& $-100^\circ\rightarrow50^\circ$& $200^\circ\rightarrow260^\circ$& $|\delta|>10^\circ$&$-120^\circ\ra30^\circ$&$30^\circ\ra78^\circ$;&$162^\circ\ra214^\circ$\\
& & & & & &$162^\circ\ra214^\circ$&\\
$T_{\rm rec}$ & \multicolumn{3}{c|}{\dotfill23\dotfill} & \multicolumn{2}{c|}{\dotfill23\dotfill}& \multicolumn{2}{c|}{\dotfill30\dotfill} \\
$\eta$ & \multicolumn{3}{c|}{\dotfill1.5\dotfill} & \multicolumn{2}{c|}{\dotfill1.07\dotfill}&  \multicolumn{2}{c|}{\dotfill1.67\dotfill} \\
$G$ (Jy/K) &  \multicolumn{3}{c|}{\dotfill0.64\dotfill} &\multicolumn{2}{c|}{\dotfill0.64\dotfill}& \multicolumn{2}{c|}{\dotfill8.514\dotfill} \\
$m$ & 7 & 7 & 7 & 10 & 10 & 7 & 7 \\
$\Omega$ (deg$^2$)$^\dagger$& \multicolumn{3}{c|}{\dotfill0.043\dotfill} & \multicolumn{2}{c|}{\dotfill0.043\dotfill}& \multicolumn{2}{c|}{\dotfill0.0027\dotfill}\\
$f$ (MHz) & 1372.5 & 1372.5 & 1374 &\multicolumn{2}{c|}{\dotfill1352\dotfill}& \multicolumn{2}{c|}{\dotfill1440\dotfill} \\
$B$ (MHz) &\multicolumn{3}{c|}{\dotfill288\dotfill} &\multicolumn{2}{c|}{\dotfill340\dotfill}& \multicolumn{2}{c|}{\dotfill100\dotfill}\\
$b$ (MHz) & \multicolumn{3}{c|}{\dotfill3\dotfill} &\multicolumn{2}{c|}{\dotfill0.390625\dotfill}&  \multicolumn{2}{c|}{\dotfill0.390625\dotfill} \\
$t_{\rm samp}$ ($\mu$s) & 125 & 250& 125&\multicolumn{2}{c|}{\dotfill64\dotfill}& \multicolumn{2}{c|}{\dotfill64\dotfill} \\
$T_{\rm obs}$ (h) $^\dagger$ & 11,859 & 20,248 & 6,924 & 7,849 & 15,041 & 3,214 & 1,974\\
DM limit & 3000 & 5000 & 1500 & 2000 & 5000 & 1000 & 2038 \\
$\dmmw$ $\bar x, \tilde x$ & 160, 120  & 701, 600 & 232, 193 & 49, 36 & 383, 289  & 396, 190 & 180, 182 \\
$T_{\rm sky}\ \bar x, \tilde x$ (K) Ê& 1.70, 1.43 & 6.14, 5,14 & 1.21, 1.19 & 0.85, 0.70 & 3.18, 2.22 & 3.35, 1.84 & 1.31, 1.31 \\
$\taus\ \bar x, \tilde x$ ($\mu$s) & 188, 0.18 & $11.7${\sc e}$3$, 70.6 & 4.75, 3.06 & 36.2, $4.75${\sc e}$^{\operatorname{--}}3$ & 4125, 2.78 & $1.31${\sc e}$3$, 2.79 & 1.79, 1.74 \\
\hline
\hline
$[\gamma=-1.5]$& & & & & & & &  {\bf TOTAL}\\
$N_{\rm A}$, local& 1.17 & 1.61 & 0.71& (4)$^\ddagger$ & 6.81 & 1.19 & 0.81 & {\bf 12.3}\\
$N_{\rm A}$, exgal& 1.08 & 1.07 & 0.61 & (4)$^\ddagger$& 5.46 & 0.90 & 0.73 & {\bf 9.85}\\ 
\hline
\textbf{\textit{N}}$_{\rm A,}$ actual& {\bf 1}  & {\bf 0} & {\bf 0}  & {\bf (4)}$^\ddagger$ & {\bf 0}  & {\bf 0}  & {\bf 1} & \textbf{2}\\
\hline
\end{tabular}\\
$^\dagger$ Note that we have quoted, and use in computations, the single-beam field of view for the 7- and 13-beam PALFA and Parkes receivers; the total observation time reflects this accordingly. Numerical results are given for $\gamma=-1.5$ (see also Fig.\,\ref{fig:nums}).\\
$^\ddagger$ These values were used as reference values (Eq.\,\ref{eq:rates}) to make the prediction values for the other surveys, therefore are not included in the total sum.
}
\end{table*}


\begin{figure}
\centering
\includegraphics[width=0.47\textwidth,trim=22mm 22mm 19mm 22mm, clip]{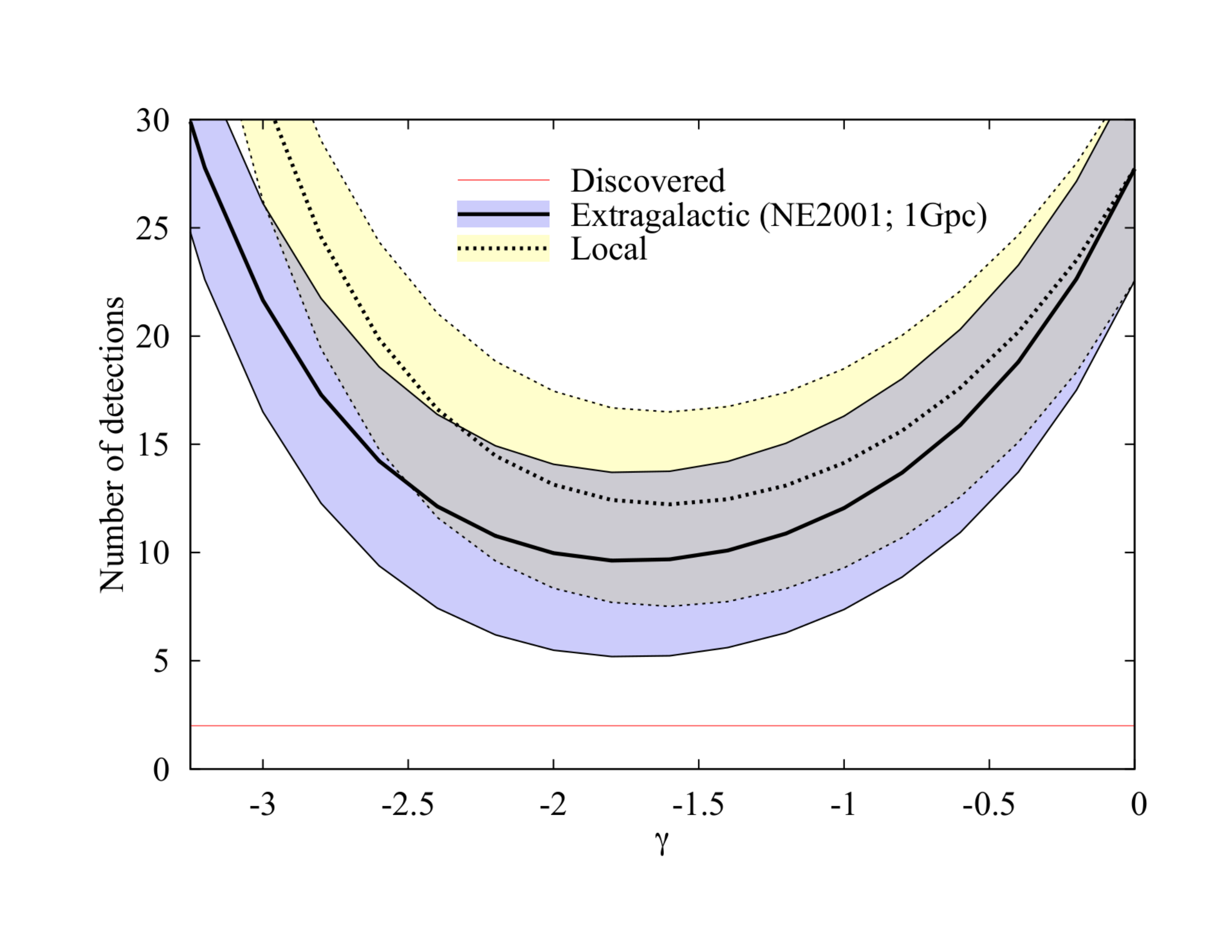}
\caption{A comparison of the single detection in the $|gb|<30^\circ$ archival surveys with that predicted for an extragalactic FRB population at an average detected distance of 1\,Gpc, and for a local population, against the pulse energy distribution parameter $\gamma$. We base prediction values on the four high-latitude detections of \citet{thornton13}. The error ranges represent a 95\% confidence interval on the predicted values. The assumption of 1\,Gpc average FRB distance does not strongly impact the results presented in this Figure; see \S\ref{sec:cosmo} and \S\ref{sec:relrates} for a discussion of this.
}\label{fig:nums}
\end{figure}

\subsection{Relative detection rates in the high-latitude HTRU and low-latitude archival surveys}\label{sec:relrates}
We now compare the detections of low-latitude archival transient searches with their expected FRB discovery rate based on the high-latitude detections of \citet{thornton13}, using the analysis of Section \ref{sec:gpos} to examine expectations for the local/isotropic and extragalactic sky distribution hypotheses.

We calculated $\Na$ using Eq.\,\ref{eq:na} with the average properties of the \citet{thornton13} survey (i.\,e.~DM$_2=\langle\dmmw\rangle+770\,\dmunits=819\,\dmunits$; $\langle T_{\rm sky2}\rangle=0.85\,$K; $\langle\tau_{{\rm s}2}\rangle=36.2\,\mu$s,
as reported in Table\,\ref{table:sparams}).
We report $\Nt\sum_p\theta(p)$ for each survey and the net $\Na$ value in the bottom rows of Table\,\ref{table:sparams}, using local and extragalactic factors to calculate $\theta(p)$. Figure\,\ref{fig:nums} demonstrates the $\Na$ expected for a range of $\gamma$ values.

Only one burst has been detected in the archival surveys we analyze. To understand the statistical significance of the difference between this and the expectations reported in Table\,\ref{table:sparams}, we take the probability of getting $\Na(=1)$ events given that $\Nt(=4)$, which has a Poissonian distribution. 

First we test the local/isotropic distribution hypothesis. As seen in Fig.\,\ref{fig:nums}, the prediction is closest to our detected value at its minimum at $\gamma\simeq-1.6$. At that point, we reject a uniform sky distribution at a probability of $P=0.00037$, which corresponds to the equivalent of a confidence of $\sim$3.6$\sigma$. In other words, even after accounting for the known effects of $T_{\rm sky}$ and instrumentation sensitivity, the observed burst rate's sky distribution is non-uniform to high significance, with a deficit at low galactic latitudes, for any value of $\gamma$.

In our extragalactic origin hypothesis, accounting for {\sc ne2001} predictions for scattering and dispersion at low Galactic latitudes helps to reconcile the low-latitude deficit, however still does not provide a confident agreement with the archival surveys' singular discovery. With our extragalactic formulation, $\Na$ is at a minimum at $\gamma=-1.7$. The probability of detecting one burst at that $\gamma$ is $P=0.0031$, while the probability at the standard Euclidean $\gamma=-1.5$ is $P=0.0025$, reflecting a disagreement at a confidence of around 3$\sigma$. This is not sufficient to rule out that FRBs are extragalactic,  particularly considering our simplifying assumptions about the Galaxy and the FRB population. To resolve the discrepancy here, the extragalactic sky-distribution model would have to account for even fewer bursts detected in the low-latitude archival surveys. We review our assumptions here, along with how they might change the predicted numbers, and a note on which might account for the observed discrepancy:
\begin{itemize}
\item The average detectable FRB distance could be much closer than 1\,Gpc. As previously noted, a very nearby FRB population would imply heightened Galactic scattering, decreasing the predicted number of detections at intermediate Galactic latitudes. We have investigated this possibility, and find that FRBs would have to be at an average distance of $<$100\,Mpc to change the $\Na$ predictions in the extragalactic model by more than 5\%.
\item We used the {\sc ne2001} model to predict Galactic scattering and dispersion effects. Additional Galactic scattering or other pulse flux dampening that is not correctly modelled by {\sc ne2001} may be occurring in the Galactic plane or halo. If such effects have a large gradient across $gl$ and $gb$, they would have a major impact on our predictions. {\sc ne2001} is currently seen as the most accurate Galactic model, however its errors increase at larger $|gb|$. Subsequent publications have argued that the scale height of the ``thick disk'' in {\sc ne2001} should be larger \citep{gaensler08,schnitzeler12}. Such an alteration of the Galaxy model would raise $\dmmw$ at intermediate latitudes, and significantly ease the observed discrepancy for the extragalactic population predictions in our analysis.
\item We assumed that FRB pulses are distributed according to a power-law flux distribution ($N\propto S^{\gamma}$). If FRB distributions are non-Euclidean (i.\,e.~they are truly cosmological, and/or the progenitor population evolves significantly at $z<1$), this may account for the fewer detections at the lower Galactic latitudes, where large $T_{\rm sky}$ contribution curtails sensitivity to faint pulses.
\item We have assumed that typical FRBs have negligible intrinsic pulse widths and extragalactic scattering when compared to instrumental effects. Non-negligible values would actually increase the expected number of discoveries in the archival surveys, therefore would not resolve the observed differences.
\item Our assumption that all surveys have been searched to a sufficient DM may not be valid. We can assess this by inspecting the mean/median ${\rm DM_{MW}}$ values for each survey in Table\,\ref{table:sparams}, when compared to the DM search limit and considering the range of FRB DMs detected thus far. We see that this is a valid assumption for all but the \citet{deneva} search, for which two of five of the known FRBs might have not been spotted in the search if they occurred in the most central galaxy regions. One additional discovery from this search at higher DM would ease, but not solve, the discrepancy with our distribution predictions. The rejection of a local population would still hold at a confidence of over 3$\sigma$.
\item If the dispersion of the \citet{keane} detection is not in fact due to the excess Galactic electrons reported by \citet{bannister14}, this helps to bring the discovered and predicted populations closer. While this is true for both extragalactic and local predictions, the extragalactic prediction still maintains a better fit to the discoveries to date.
\item We have assumed a flat spectral index at 1\,GHz. However, this will not greatly effect our results as all the surveys we analyzed were observed at similar frequencies.
\end{itemize}

Finally, one remaining possibility to explain disagreements with both of our prediction scenarios is that FRBs have a Galactic origin. However, there are strong observational arguments against this possibility \citep{kulkarni+14}, and a Galactic explanation for FRBs must account for a \emph{lower} detection rate in the Galactic plane; even very nearby Galactic sources show tendencies toward Galactic-plane excesses.

We end on one final note. We are clearly dealing with small number statistics (i.\,e.~1 and 4). Although our statistical statements account for Poissonian variance, they do not account for human error; Fig.\,\ref{fig:nums} demonstrates that if one burst was missed \emph{per low-latitude survey}, we might see complete agreement with the extragalactic model predictions, and less significance in the disagreement with a uniform sky distribution or local population. We believe we have sufficiently accounted for this issue by restricting survey thresholds to ${\rm S/N}>7$, the point at which individual pulsar pulses are reliably flagged by human inspectors in manual inspection plots (to a $\sim$99\% success rate). It is furthermore possible to miss FRB events due to radio interference that can render an entire pointing unusable. However, the occurrence rate of this effect is consistent between surveys and so cannot cause the variance with Galactic position that we have demonstrated here.

%
%
  
  
\section{Summary}
We have presented the discovery of a $|gb|=20^{\circ}$ Fast Radio Burst, \src, in a search of the intermediate-Galactic-latitude surveys of \citet{ED} and \citet{BJ}. As evidenced by its flux, few-milliseconds timescale, temporal isolation, a frequency sweep that closely adheres to the cold plasma dispersion relation, and its $\sim$7-times excess DM over that expected from the Galaxy, the burst appears to be of the same population as the bursts reported by \citet{thornton13}. However, the detection rate in our search is a factor of $\sim$5 less than that of \citet{thornton13}.

This difference led us to inspect the reason for the lack of FRB discoveries in a number of low-latitude $f\simeq1$\,GHz surveys as a means to test the sky distribution of FRBs against expectations for local and extragalactic FRB origins. Based on the survey design and discoveries of \citet{thornton13}, we made predictions for the number of FRBs that should be discovered in the low-latitude archival surveys. 
Instrumental, sky temperature, and other Galactic filtering effects were considered (Eq.\,\ref{eq:rates}).

A test of FRBs as a local/solar-system origin population (i.\,e. uniform sky distribution w.r.t.\ Galactic position) revealed a rejection of this hypothesis at a confidence of 3.6$\sigma$. That is, we see a significant drop in the detection rate at low Galactic latitudes, which is strong evidence of a non-local (Galactic or extragalactic) origin.

A test of FRBs as an extragalactic population, using the {\sc ne2001} model to account for excess dispersion and scattering in the Galaxy, showed a closer agreement with discovery rates however still demonstrated a discrepancy at the $\sim$3$\sigma$ level. An inspection of our simplifying assumptions revealed several points that could produce fewer expected detections at $|gb|<30^\circ$, and thus give a closer agreement with an extragalactic FRB model: most prominently, 1) an FRB population at $<$100\,Mpc distance; 2) excessive dispersion or scattering effects in the Galaxy/halo that are not accounted for by the {\sc ne2001} model, \eg\ a larger disk scale-height.

We have not ruled out the possibility of a Galactic origin for FRBs, which could produce more complex sky-dependent detection effects than were considered here. \citet{kulkarni+14} provides an overview of several potential Galactic and extragalactic FRB progenitor models, and a look at the credibility of these possibilities.

These results must be interpreted with some caution: the small total number of FRB detections thus far, and their apparent rarity, make numerical prediction comparisons sensitive to human error during the manual inspection stage of candidate identification, and other misestimations of sensitivity thresholds. However, we believe we have suitably accounted for these effects in our analysis.

Ongoing and future surveys, particularly those which focus on Galactic latitudes $|b|<15^\circ$, will place the strong constraints on the latitude dependence of FRBs. Based on our framework, we expect 3.8 and 6.8 discoveries in the HTRU low latitude survey for extragalactic and local populations, respectively. For a survey of the inner Galaxy with the PALFA system, 1.5 and 2.8 bursts are expected from an extragalactic and local population, respectively, for a 1000\,h survey. However, the predictions for extragalactic event rates are likely to actually be lower given our results and when considering the simplifying assumptions made in our predictions (summarized in \S\ref{sec:relrates}).

Finally, we would like to make the general note that regardless of FRB origins, future searches for FRBs will apparently find greater success if they survey at Galactic latitudes $|gb|\gtrsim20$. The expression given in Eq.\,\ref{eq:rates} provides a convenient estimator for various survey design parameters to optimize FRB detection.

\section{Acknowledgements}
The Australia Telescope Compact Array Parkes Radio Telescope is part of the Australia Telescope National Facility, which is funded by the Commonwealth of Australia for operation as a National Facility managed by CSIRO. Processing for our archival search was performed on the Swinburne University Green Machine; we acknowledge the use of this supercomputer facility in this work.
We would like to thank thank B.~Stappers, D.~Thornton, J.~Deneva, and A.~Brazier for providing information about survey pointing directions which were used in this paper's analysis. We thank an anonymous referee for contributing valuable comments on this paper. Cosmological redshift calculations used the online calculator provided by \citet{cosmocalc}.


\bibliographystyle{apj}
\bibliography{bsb_emulateapj}

\end{document}